\documentclass[twocolumn,twoside,slac_two,author-year]{revtex4}
\usepackage{graphicx}
\usepackage{fancyhdr}
\usepackage[caption=false]{caption}
\usepackage[font=footnotesize]{subfig} 
\usepackage{color}

\newcommand{\fermi}{\textit{Fermi}}%
\newcommand{\FGST}{\textit{Fermi Gamma-ray Space Telescope}}%
\newcommand{\COSB}{\textit{COS-B}}%
\makeatletter
\newcommand{\ion}[2]{#1$\;${\small\rmfamily\@Roman{#2}}\relax}%
\makeatother
\newcommand{\HI}{\mbox{\ion{H}{1}}}%
\newcommand{\Xco}{\mbox{$X_\mathrm{CO}$}}%
\newcommand{\Wco}{\mbox{$W_\mathrm{CO}$}}%
\newcommand{\Hmol}{\mbox{$\mathrm{H}_{2}$}}%
\newcommand{\degr}{\mbox{$^\circ$}}%
\newcommand{\piz}{\mbox{$\pi^0$}}%
\newcommand{\xunit}{\mbox{$\times10^{20}\ \mathrm{cm}^{-2}(\mathrm{K\ km\ s}^{-1})^{-1}$}}%
\newcommand{\MeV}{\mbox{$\mathrm{MeV}$}}%
\newcommand{\GeV}{\mbox{$\mathrm{GeV}$}}%
\newcommand{\prel}{\textit{PRELIMINARY}}%
\newcommand{\kpc}{\mbox{$\mathrm{kpc}$}}%
\newcommand{\pc}{\mbox{$\mathrm{pc}$}}%
\newcommand{\sun}{\odot}%
\newcommand{\stat}{\mbox{$\mathrm{stat}$}}%

\begin{document}

\title{Diffuse Gamma-ray Observations of the Orion Molecular Clouds}

%

\author{Akira Okumura}
\email{oxon@juno.phys.s.u-tokyo.ac.jp}
\affiliation{Department of Physics, University of Tokyo, 7-3-1 Hongo, Bunkyo-ku, Tokyo, 113-0033, Japan}
\author{Tune Kamae}
\email{kamae@slac.stanford.edu}
\affiliation{SLAC National Accelerator Laboratory, Stanford, CA 94025, USA}
\author{on behalf of the \fermi/LAT Collaboration}

\begin{abstract}
We report on a preliminary analysis of the diffuse gamma-ray observations of local giant molecular clouds Orion A and B with the Large Area Telescope onboard the \FGST. The gamma-ray emission of the clouds is well explained by hadronic and electromagnetic interactions between cosmic rays and nuclei in the clouds. Consequently, we obtain the total masses of the Orion A and B clouds to be $(80.6\pm7.5\pm 4.8)\times 10^3 M_\sun$ and $(39.5\pm5.2\pm2.6)\times 10^3 M_\sun$, respectively, for the distance to the clouds of $400\ \pc$ and the Galactic CR spectrum predicted by GALPROP on the local observations of CRs. The structure of molecular clouds has been extensively studied by radio telescopes, especially using the line intensity of CO molecules (\Wco) and a constant conversion factor from \Wco\ to $N(\Hmol)$ ($\equiv\Xco$). However, this factor is found to be significantly different for Orion A and B: $1.76\pm0.04\pm0.02$ and $1.27\pm0.06\pm0.01$, respectively.

\end{abstract}

\maketitle

\thispagestyle{fancy}


\section{INTRODUCTION}

Diffuse emission of $> 100\ \MeV$ gamma rays in the Galaxy is mainly induced by hadronic interactions between the Galactic cosmic rays (CRs) and interstellar medium (ISM), via the production of \piz\ particles and their subsequent decay into photons. This emission can be used to study CRs and the structure of ISM, because its emission can be written as the product of the CR flux and the density of ISM, and because the gamma-ray spectral shape preserves the CR properties. Observations of the emission have mainly two unique advantages. First, we can study CRs at the location of the gamma-ray emission, while direct measurements of CRs at the Earth cannot extract their positional information which is already lost during propagation in the interstellar magnetic fields of the Galaxy. Second, \piz\ emission is not affected by the gas condition, such as the temperature of interstellar dust, and the CO to \Hmol\ ratio which are often used to estimate the column density of ISM.

Since the early stage of gamma-ray astronomy, diffuse emission from \HI\ gas and molecular clouds (MCs) has been extensively studied to understand Galactic CRs and the ISM (e.g. \cite{Kraushaar:1972:High-Energy-Cosmic-Gamma-Ray-O,Hunter:1997:EGRET-Observations-of-the-Diff}). Among them, the Orion A and the Orion B MCs are two of the best targets, and have been well studied by pioneering gamma-ray telescopes \cite{Caraveo:1980:COS-B-observation-of-high-ener,Bloemen:1984:Gamma-rays-from-atomic-and-mol,Digel:1995:EGRET-Observations-of-Gamma-ra,Digel:1999:EGRET-Observations-of-the-Diff}. This is because they are considered to be the archetypes of giant MCs located near the Earth, and have been surveyed by radio telescopes especially using the line emission of CO molecules (e.g. \cite{Maddalena:1986:The-Large-System-of-Molecular-,Wilson:2005:A-uniform-CO-survey-of-the-mol}). From the analysis side, their gamma-ray fluxes are strong enough to be distinguished from other diffuse emission components (\HI\ gas, inverse Compton, extragalactic diffuse emission), because they located about only $\sim400$~pc away from the Sun, their total mass is of the order of $10^5M_\odot$, and their Galactic coordinates $(\ell\simeq 210\degr, b\simeq-15\degr)$ are far from the Galactic plane and the Galactic center.

In the study of MCs, the biggest difficulty is that \Hmol, the main component of MCs, cannot be observed directly. Hence, CO molecules which are the second abundant molecules, have been widely used as a tracer of \Hmol\ distribution with a conventional factor $\Xco\equiv N(\Hmol)/\Wco$ which converts velocity integrated CO line intensity, \Wco\ to \Hmol\ column density, $N(\Hmol)$. For example, $\Xco = (1.8\pm0.3) \xunit$ was derived by comparing a CO survey with \HI\ and dust observations over a Galactic scale \cite{Dame:2001:The-Milky-Way-in-Molecular-Clo}. This factor has been also determined by diffuse gamma-ray observations, e.g. $(2.3\pm0.3)\xunit$ by \COSB \cite{Strong:1988:The-radial-distribution-of-gal}, and $(1.35\pm0.15)\xunit$ for the Orion region by EGRET \cite{Digel:1999:EGRET-Observations-of-the-Diff}. However, due to the limited angular resolution and photon statistics of the preceding telescopes, they could not study \Xco\ for individual clouds. Besides, a mystery of ``GeV excess'' which shows a disagreement between the observed diffuse gamma-ray spectrum and the local CR spectra, was reported by the EGRET \cite{Hunter:1997:EGRET-Observations-of-the-Diff}.

Since the EGRET era, much progress has been made in gamma-ray observations and studies on the Orion A/B clouds. The most important progress is the launch of the Large Area Telescope (LAT) onboard the \FGST\ (\fermi) \cite{Atwood:2009:The-Large-Area-Telescope-on-th}. Its large effective area ($\sim1\ \mathrm{m}^2$) and wide energy band ($20\ \MeV$ to $>300\ \GeV$) make it possible to study \piz\ gamma rays with large photon statistics above $1\ \mathrm{GeV}$. In addition, the energy-dependent LAT angular resolution is a few times better than that of EGRET. As a result, we are able to resolve the structure of the Orion clouds on a scale of $\sim1\ \degr$ using higher-energy photons of better angular resolutions, and also able to compare the gamma-ray energy spectrum with the predicted one based on the CR spectra observed at the Earth. \piz\ gamma-ray emissivity of hadronic interactions was modeled in several articles using recent results of accelerators and theoretical studies (e.g. \cite{Kamae:2006:Parameterization-of-gamma-epm-,Mori:2009:Nuclear-enhancement-factor-in-,Gaisser:1992:Cosmic-Ray-Secondary-Antiproto}), so that the mass of the clouds can be calculated backwards from the gamma-ray emission and the distance to the clouds which was measured accurately from recent MASER observations of the Orion nebula in the Orion A \cite{Hirota:2007:Distance-to-Orion-KL-Measured-,Sandstrom:2007:A-Parallactic-Distance-of-389-,Menten:2007:The-distance-to-the-Orion-Nebu,Kim:2008:SiO-Maser-Observations-toward-}.

\section{OBSERVATIONS AND DATA}

The data used in this analysis were obtained in the nominal all-sky survey mode of the LAT between 4 August 2008 and 15 August 2009. Among all the LAT events, we selected ones classified as \textit{P6\_V3\_DIFFUSE} class \cite{Atwood:2009:The-Large-Area-Telescope-on-th}. The reconstructed energy range and zenith angle were limited to $200\ \MeV$ -- $20\ \GeV$, and $<105\degr$, respectively. Gamma rays in a circular region of radius $20\degr$ centered at $(\ell, b) = (211\degr, -17\degr)$ were then selected for later analyses.

\begin{figure}
\includegraphics[width=80mm]{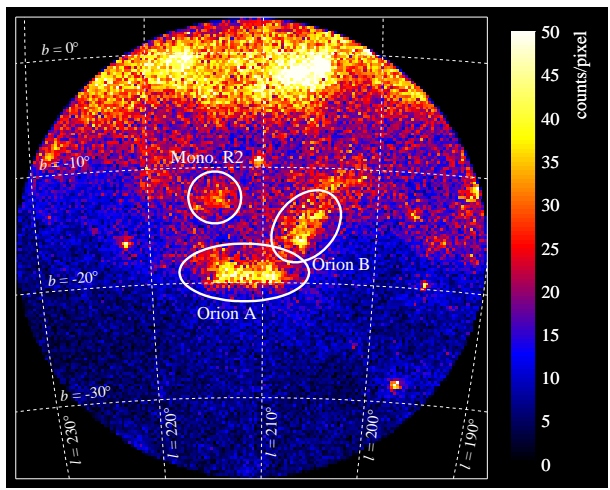}
\caption{Gamma-ray ($200$~MeV to $20$~GeV) count map of the Orion region binned in $0.25\degr\times0.25\degr$ pixels with Hammer-Aitoff projection. The coordinate center and the radius of the region are ($\ell, b) = (211\degr, -17\degr)$ and $20\degr$, respectively. In addition to the Galactic plane, three molecular clouds, Orion A, Orion B and Monoceros R2 are visible as bright extended sources.}
\label{fig_countmap}
\end{figure}

All events in the selected data are binned in $160\times160$ equal area pixels with the Hammer-Aitoff projection as shown in Fig.~\ref{fig_countmap}. The Orion A and B are easily recognized by their count excess from the surrounding diffuse emission and point sources.

The reconstructed energies of the events are binned in a logarithmic series of $16$ between $200\ \MeV$ and $20\ \GeV$. The LAT exposure of each energy bin was calculated from the LAT pointing history and instrumental response function (IRF) using ScienceTools v9r15p4.

\section{ANALYSIS}

\subsection{EXTRACTION OF GAMMA-RAY EMISSION FROM ORION A AND B}
\label{subsec_extraction}
To study the gamma-ray emission associated with the Orion MCs (molecular gas), other emission components must be subtracted from Fig.~\ref{fig_countmap}. The emission in the region consists of several components. The dominant ones are diffuse \piz\ gamma-ray and electron bremsstrahlung emission induced by interactions between CRs and \HI/\Hmol\ gas. In addition, inverse Compton (IC) scattering of CR electrons off interstellar radiation fields exits in the region. Extragalactic diffuse emission, and residual instrumental background are classified as diffuse emission. We refer to the latter two components as ``isotropic component''. The LAT detection of some point sources in the region has been reported. However, there is no intense point source overlapping the Orion MCs \cite{Abdo:2009:Fermi/Large-Area-Telescope-Bri}.

\begin{figure*}
  \centerline{
    \subfloat[Extracted gamma-ray intensity]{
      \includegraphics[width=80mm]{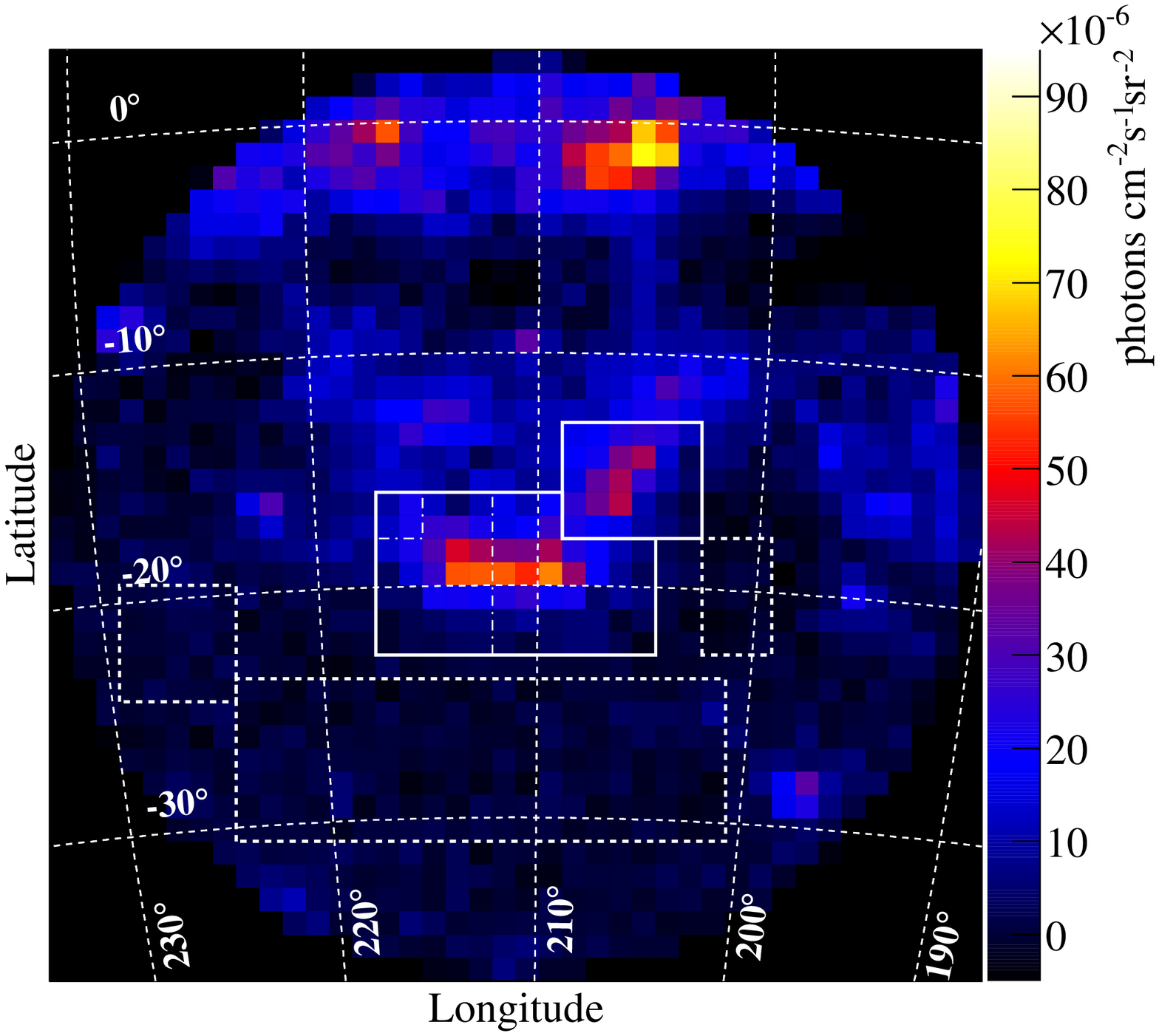}
      \label{fig_intensity_map}
    }
    \hfil
    \subfloat[Modeled \Wco-based intensity]{
      \includegraphics[width=80mm]{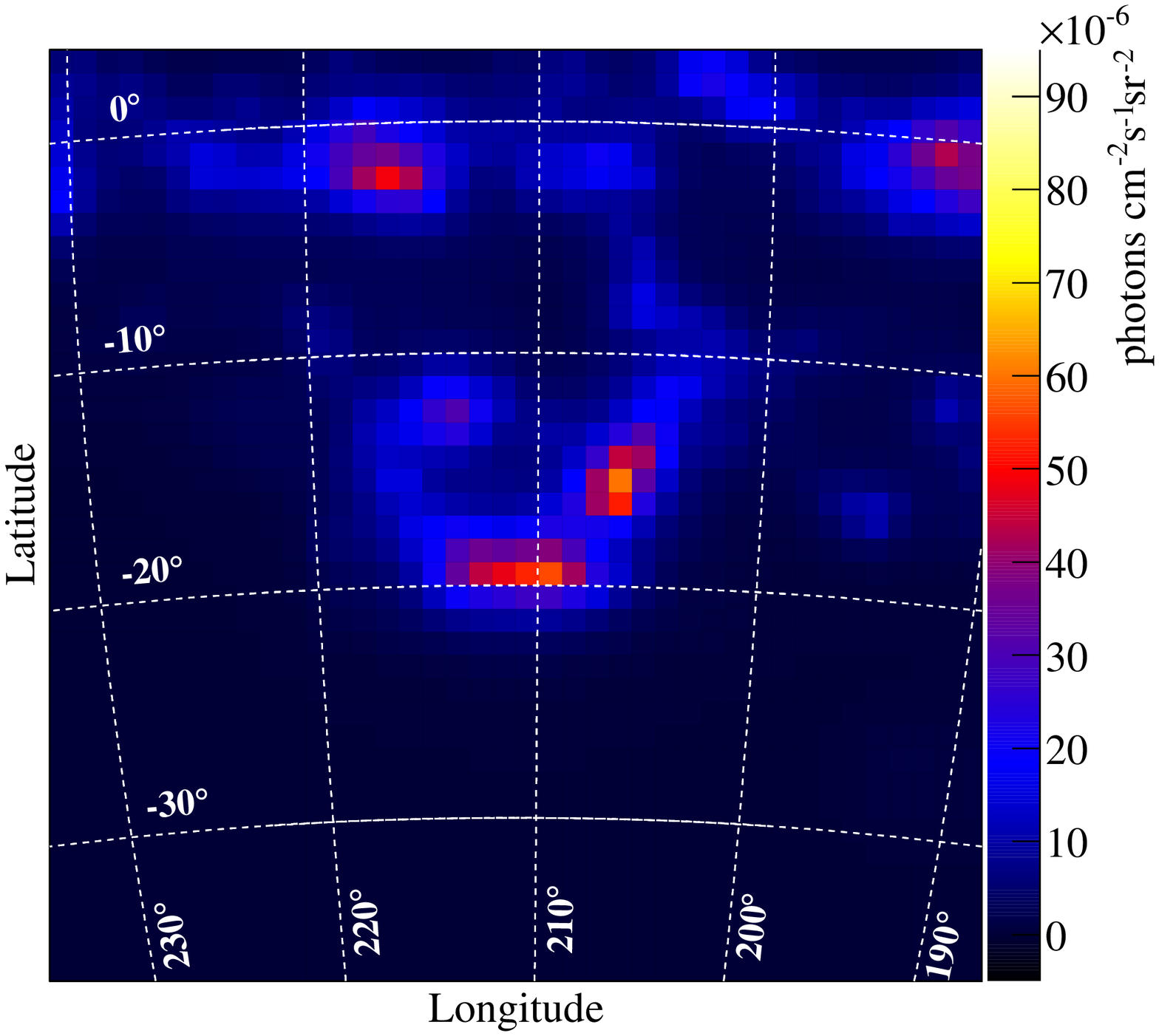}
      \label{fig_CO}
    }
  }
  \caption{(a) Same as Fig. \ref{fig_countmap}, but shown after subtracting the other diffuse components. The unit is integrated photon intensity between $200\ \MeV$ and $20\ \GeV$. Solid lines show the definitions of the regions of Orion A and B. Rectangles with dashed lines show the regions which are used to estimate the isotropic component. A straight dot-dashed line at $\ell=212\degr$ is the boundary of two separated regions in Orion A. (b) A modeled gamma-ray intensity map simulated from a constant $\Xco=1.5\xunit$, \Wco\ map \cite{Dame:2001:The-Milky-Way-in-Molecular-Clo}, the LAT response, the emissivity of \piz\ gamma and electron bremsstrahlung, and the CR flux predicted by GALPROP. (These figures are still \prel)
}
\end{figure*}

In order to extract the Orion A and B, we modeled other emission components using a computer program called GALPROP which calculates CR flux and diffuse gamma-ray emission in the Galaxy  \cite{Strong:1998:Propagation-of-Cosmic-Ray-Nucl,Strong:2000:Diffuse-Continuum-Gamma-Rays-f}. It uses the data of \HI\ and CO surveys to calculate the target mass of CR interaction \cite{Kalberla:2005:The-Leiden/Argentine/Bonn-LAB-,Dame:2001:The-Milky-Way-in-Molecular-Clo}\footnote{We assumed a constant spin temperature of $125$~K for the \HI\ map.}. Its calculation is known to be consistent with the LAT gamma-ray observations in a galactic scale except for the Galactic plane, when reasonable input parameters based on many observations and experiments are given \cite{Abdo:2009:Fermi-Large-Area-Telescope-Mea,Strong:2009:Large-scale-Galactic-diffuse-g,Abdo:2009:Fermi-LAT-Observation-of-Diffu}. Each set of input parameters is referred to as a GALDEF file. In this analysis, we use GALDEF 54\_77Xvarh7S in which CR proton and electron fluxes are scaled by factors of 1.15 and 1.75, respectively, to reproduce the LAT observations.

Before we subtracted the \HI\ contribution in the region, its normalization was estimated in nearby two $10\degr$ radius regions near the Orion MCs where no known large MCs exists. The regions are centered at $(\ell, b) = (230\degr, -16\degr)$ and $(210\degr, -32\degr)$, and the obtained fluxes are $1.15$ and $0.98$ times those predicted by GALPROP, respectively. Thus we multiply 1.07 to the GALPROP prediction for gamma-ray intensity from \HI\ gas, and add $8\%$ to the systematic uncertainty of \HI\ subtraction process. The scaled \HI\ emission was convolved with the LAT IRF, and subtracted from the pixelized data in each energy bin. In addition to this, IC emission was also subtracted. Since the IC contribution is smaller than other emission, its uncertainty does not affect our results.

Finally, the isotropic component was estimated from surrounding regions (background regions) around the Orion MCs, then subtracted from the region-of-interest. Fig.~\ref{fig_intensity_map} shows the gamma-ray intensity map after the subtraction method described above . It was divided by the LAT exposure, and rebinned in $1\degr\times1\degr$ pixels. We defined the boundary of Orion A and B with solid lines in Fig.~\ref{fig_intensity_map}.

\subsection{ENERGY SPECTRA AND TOTAL MASSES OF THE CLOUDS}

\begin{figure*}
  \centering
  \includegraphics[width=120mm]{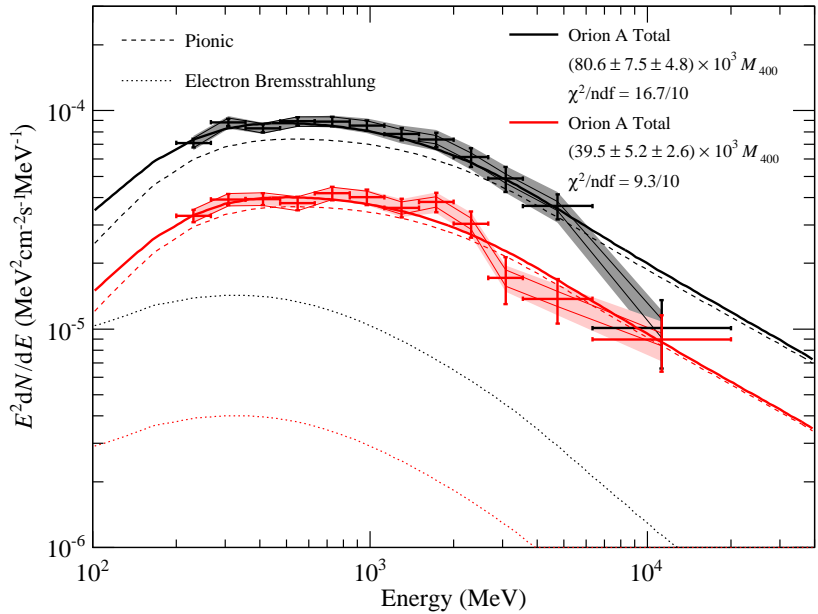}
  \caption{Energy spectra of the Orion A (black) and B (red) fitted with modeled spectra of \piz\ gamma (dashed) and electron bremsstrahlung (dotted). Statistical errors are shown with bars, and the systematic errors of the LAT response are shown with shaded area. Polygons of solid lines show the systematic uncertainties of \HI\ subtraction process. (This figure is still \prel)}
  \label{fig_spectra}
\end{figure*}

The energy spectra associated with the Orion A and B clouds are shown in Fig.~\ref{fig_spectra}. All photons in the boundary regions in Fig.~\ref{fig_intensity_map} are collected for each cloud. They are fitted with \piz\ and bremsstrahlung components with two free normalization parameters. Here, we calculated \piz\ gamma-ray emissivity using a recent parameterized model \cite{Kamae:2006:Parameterization-of-gamma-epm-}. Input CR spectra at the Orion region\footnote{Corresponding to a cylindrical Galactic location $(R = 8.5\ \kpc, Z=-0.14\ \kpc)$.} predicted by GALPROP are assumed, which are $\sim8\%$ smaller than the observed CR flux at the Earth \cite{Sanuki:2000:Precise-Measurement-of-Cosmic-,Alcaraz:2000:Protons-in-near-earth-orbit,Alcaraz:2000:Helium-in-near-Earth-Orbit}. Gamma-ray inclusive cross sections for proton-He, alpha-H, alpha-He, and heavier CR metals are scaled using the method described in \cite{Mori:1997:The-Galactic-Diffuse-Gamma-Ray,Gaisser:1992:Cosmic-Ray-Secondary-Antiproto}. A factor 1.02 as a contribution from ISM metal was finally multiplied to the calculated \piz\ emissivity \cite{Mori:2009:Nuclear-enhancement-factor-in-}. The bremsstrahlung spectrum was calculated using GALPROP.

The obtained fit results show good agreements with the \piz\ spectrum. This means that gamma-ray emission from MCs is dominated by known physical processes, i.e. interactions between CRs and the gas. Therefore, we are able to calculate the total mass of the clouds. Assuming the distance to the Orion clouds to be $400\ \pc$ \cite{Hirota:2007:Distance-to-Orion-KL-Measured-,Sandstrom:2007:A-Parallactic-Distance-of-389-,Menten:2007:The-distance-to-the-Orion-Nebu,Kim:2008:SiO-Maser-Observations-toward-}, the mass of the Orion A and B are estimated to be $(80.6 \pm 7.5_{(\stat)}\pm 4.8_{(\HI)})\times 10^3 M_\sun$ and $(39.5 \pm 5.2_{(\stat)} \pm 2.6_{(\HI)})\times 10^3 M_\sun$, respectively. The second terms indicate statistical errors of the fit, and the third ones are systematic errors introduced by the uncertainty of \HI\ subtraction explained in subsection~\ref{subsec_extraction}.

\subsection{CORRELATION BETWEEN GAMMA-RAY AND CO INTENSITY}

Assumption of a constant \Xco\ in small scales ($\sim10\ \pc$ to $\sim100\ \pc$) or Galactic scale has been widely used in studies of MCs (e.g. \cite{Dame:2001:The-Milky-Way-in-Molecular-Clo,Digel:1999:EGRET-Observations-of-the-Diff}), while gradient of \Xco\ is also discussed \cite{Arimoto:1996:CO-to-H2-Conversion-Factor-in-,Strong:2004:The-distribution-of-cosmic-ray}. However, there is room to reconsider this simple assumption that \Wco\ can be used to trace the structure of MCs \cite{Boulanger:1998:CO-and-IRAS-observations-of-th,Greiner:2005:Unveiling-Extensive-Clouds-of-}.

Utilizing the good angular resolution and large photon statistics of the LAT, the correlation between gamma-ray intensity and a \Wco\ map can be studied in a scale of $\sim1\degr$. Fig.~\ref{fig_CO} shows a modeled gamma-ray intensity map based on a CO survey and GALPROP calculation with a constant \Xco\ of $1.5\ \xunit$, where the LAT IRF is convolved.

\begin{figure*}
  \centering
  \includegraphics[width=135mm]{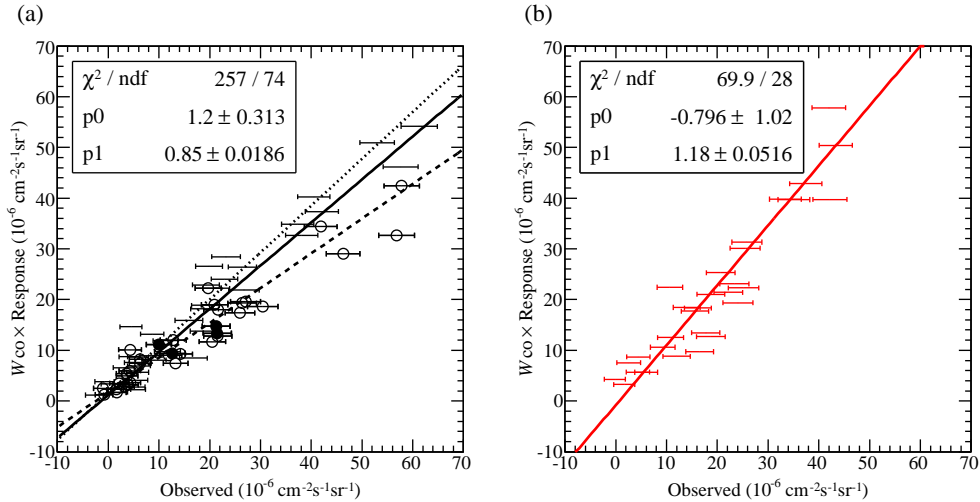}
  \caption{Correlations between the measured gamma-ray intensity ($x$) and a model prediction by \Wco\ ($y$) assuming $\Xco=1.5 \xunit$. Data points are fitted with a linear function $y=p_0 + p_1 x$. (a) Orion A. Circles correspond to the data in pixels of $\ell>212\degr$. (b) Orion B. (This figures are still \prel)}
  \label{fig_correlation}
\end{figure*}

Fig.~\ref{fig_correlation} shows the pixel-by-pixel correlation between the observed gamma-ray intensity (Fig.~\ref{fig_intensity_map}; $x$) and the \Wco\ model (Fig.~\ref{fig_CO}; $y$). The best fit results by a linear function ($y = p_0 + p_1x$) are also shown. If the gamma-ray intensity can be presented by the product of a constant CR flux and a constant \Xco, the slopes of the best-fit functions become constant. However, we found that the obtained slopes are significantly different for the Orion A and B clouds, while these clouds are thought to be in a common environment since their birth. In Fig.~\ref{fig_correlation}a, two additional linear functions are shown: one is fitted with the larger longitude region in Orion A ($\ell>212\degr$), and the other is for the smaller longitude region ($212\degr>\ell$) (see Fig.~\ref{fig_intensity_map}).

Since we modeled the vertical values using a constant \Xco\ of $1.5\ \xunit$, the corresponding \Xco\ of Orion A and B are $(1.76\pm0.04_{(\stat)}\pm0.02_{(\HI)})\times10^{20}$ and $(1.27\pm0.06_{(\stat)}\pm0.01_{(\HI)})\times10^{20}$, respectively.

\section{DISCUSSION}

We obtained the energy spectra of the Orion A and B, and showed that they can be explained by the CR interactions with the nuclei in the gas. Thus, a linear correlation between gamma-ray intensity and the column density of the clouds are expected, because gamma-ray emission is not affected by environmental condition of the gas. In fact, the correlation between gamma-ray intensity and the \Wco-based modeled map holds a linearity for roughly one decade with only a few 10\% deviations. This implies that cosmic rays of energy above $\sim1\ \GeV$ can penetrate dense cores of molecular clouds. However, the correlation slopes in the Orion region was found to be not constant as shown in Fig.~\ref{fig_correlation}.

There are some possible interpretations of the different correlation slopes. In the analysis, we assumed a constant CR flux in the Orion region. However, if it is significantly different in the three separated regions in Fig.~\ref{fig_intensity_map}, the gamma-ray intensity also varies according to the CR flux variation which might be caused by the strong magnetic field in molecular clouds.

 On the contrary, if the CR flux is almost constant in the region, we need to consider nonuniformity of \Xco\ in the region. While the CO line ($J=1-0$) is the de fact standard of mass tracers of molecular clouds, the CO to \Hmol\ ratio can be varied by the condition of each cloud. By comparing \Wco\ maps with dust observations or gamma-ray observations \cite{Boulanger:1998:CO-and-IRAS-observations-of-th,Greiner:2005:Unveiling-Extensive-Clouds-of-}, it is known that there exists gas which is not traced by CO or \HI\ observations. The nonuniformity of \Xco\ in the Orion region can also be explained by this. In fact, IR emission from interstellar dust and visual extinction by dust are stronger in the left half of Orion A than that expected from CO observations \cite{Schlegel:1998:Maps-of-Dust-Infrared-Emission,Dobashi:2005:Atlas-and-Catalog-of-Dark-Clou}. Therefore, if there exist \Hmol\ molecules not fully traced by \Wco, but traced by gamma-ray and dust, the \Xco\ variation shown in Fig.~\ref{fig_correlation} can be explained. In addition to \Hmol, it is possible that a part of \HI\ gas was not subtracted adequately, because its spin temperature is relatively low in cold molecular clouds compared to surrounding diffuse \HI\ region. While we assumed a constant spin temperature of $125$~K in the region, optically thick \HI\ gas whose temperature is low may not have been corrected properly and contribute to the gamma-ray emission, especially in the left part of Orion A. 

While the gamma-ray emission from \HI\ and \Hmol\ cannot be distinguished from each other, our observations revealed non-linear relation between gamma-ray and \Wco\ maps, which will enable us to understand the nature of molecular clouds in detail, in addition to the Galactic CRs. More photon statistics in future is expected to show the smaller structure of the clouds.

\bigskip 
\begin{acknowledgments}
The \textit{Fermi} LAT Collaboration acknowledges generous ongoing support from a number of agencies and institutes that have supported both the development and the operation of the LAT as well as scientific data analysis. These include the National Aeronautics and Space Administration and the Department of Energy in the United States, the Commissariat \`a l'Energie Atomique and the Centre National de la Recherche Scientifique / Institut National de Physique Nucl\'eaire et de Physique des Particules in France, the Agenzia Spaziale Italiana and the Istituto Nazionale di Fisica Nucleare in Italy, the Ministry of Education, Culture, Sports, Science and Technology (MEXT), High Energy Accelerator Research Organization (KEK) and Japan Aerospace Exploration Agency (JAXA) in Japan, and the K.~A.~Wallenberg Foundation, the Swedish Research Council and the Swedish National Space Board in Sweden.

Additional support for science analysis during the operations phase is gratefully acknowledged from the Istituto Nazionale di Astrofisica in Italy and the Centre National d'\'Etudes Spatiales in France.
\end{acknowledgments}

\bigskip 

\color{white}

\end{document}